\documentclass[twocolumn,superscriptaddress,aps,preprintnumbers,amsmath,amssymb,prl,nofootinbib]{revtex4}

\usepackage{epsfig}
\usepackage{amsmath}
\usepackage{amssymb}
\usepackage{subfigure}
\usepackage{cancel}
\usepackage{textcomp}
\usepackage{calc}
\usepackage{graphicx}
\usepackage{dcolumn}
\usepackage{color}
\usepackage{xcolor}
\usepackage{pifont}

\usepackage{hyperref}
\hypersetup{colorlinks=true,urlcolor=blue,linkcolor=red,citecolor=green!60!black}

\begin{document}


\title{Imprint of a scalar era on the primordial spectrum of gravitational waves}

\author{Francesco D'Eramo}
\affiliation{Dipartimento di Fisica e Astronomia, Universit\`a degli Studi di Padova, Via Marzolo 8, 35131 Padova, Italy}
\affiliation{Istituto Nazionale di Fisica Nucleare (INFN), Sezione di Padova, Via Marzolo 8, 35131 Padova, Italy}

\author{Kai Schmitz}
\affiliation{Dipartimento di Fisica e Astronomia, Universit\`a degli Studi di Padova, Via Marzolo 8, 35131 Padova, Italy}
\affiliation{Istituto Nazionale di Fisica Nucleare (INFN), Sezione di Padova, Via Marzolo 8, 35131 Padova, Italy}


\begin{abstract}

Upcoming searches for the stochastic background of inflationary
gravitational waves (GWs) offer the exciting possibility to probe
the evolution of our Universe prior to Big Bang nucleosynthesis.
In this spirit, we explore the sensitivity of future GW observations
to a broad class of beyond-the-Standard-Model scenarios that lead
to a nonstandard expansion history.
We consider a new scalar field whose coherent oscillations
dominate the energy density of the Universe at very early times,
resulting in a \textit{scalar era} prior to the standard
radiation-dominated era.
The imprint of this scalar era on the primordial GW spectrum
provides a means to probe well-motivated yet elusive models
of particle physics.
Our work highlights the complementarity of future GW observatories
across the entire range of accessible frequencies.
\end{abstract}


\date{\today}
\maketitle


\noindent\textbf{Introduction.} The detection of gravitational
waves (GWs)~\cite{Abbott:2016blz} by
LIGO~\cite{Harry:2010zz,TheLIGOScientific:2014jea} and
Virgo~\cite{TheVirgo:2014hva} has rung in the age of GW astronomy.
While all signals observed thus far have been of astrophysical origin,
\textit{i.e.}, the coalescence of compact binaries~\cite{LIGOScientific:2018mvr},
it is expected that GWs will also open a window on
cosmology in the foreseeable future~\cite{Maggiore:1999vm}.
Within the next decades, a multitude of GW observatories will
go online, ranging from third-generation ground-based interferometers
over satellite-borne interferometers in space to pulsar timing arrays
monitored by new radio telescopes (see~\cite{McWilliams:2019fng,Caldwell:2019vru,
Kalogera:2019sui,Cornish:2019fee,LIGOScientific:2019vkc}
for an outlook on GW astronomy in the 2020's).
These experiments promise to be sensitive to a wealth of hypothetical
cosmological GW signals\,---\,such as the stochastic GW background (SGWB)
from inflation, GWs from phase transitions in the early
Universe, or GWs emitted by topological defects
(see~\cite{Caprini:2018mtu,Christensen:2018iqi} for recent review articles)\,---\,and
even to anisotropies in the GW signal~\cite{Geller:2018mwu}.
The direct observation of any of these cosmological signals would allow one
to chart unexplored territory that cannot be accessed 
by conventional cosmological probes such as the cosmic microwave
background (CMB) or the primordial abundances of light elements
generated during Big Bang nucleosynthesis (BBN). 


In this paper, we will be concerned with the primordial
SGWB from inflation~\cite{Grishchuk:1974ny,Starobinsky:1979ty,Rubakov:1982df}
(see~\cite{Guzzetti:2016mkm} for a review).
An intriguing feature of inflationary GWs is that they act as a logbook
of the expansion history of our Universe throughout its
evolution~\cite{Seto:2003kc,Boyle:2005se,Boyle:2007zx,
Kuroyanagi:2008ye,Nakayama:2009ce,Kuroyanagi:2013ns,Jinno:2013xqa,Saikawa:2018rcs}.
Indeed, the detailed time evolution of the Hubble rate during
the expansion determines the transfer function that describes how
gravitational waves at different frequencies are redshifted to the present day.
This property turns primordial GWs into a powerful tool that grants access to
the thermal history of our Universe prior to BBN.
Primordial GWs offer, \textit{e.g.}, an opportunity to measure
the reheating temperature after inflation~\cite{Nakayama:2008ip,Nakayama:2008wy,
Kuroyanagi:2011fy,Buchmuller:2013lra,Buchmuller:2013dja,Jinno:2014qka,Kuroyanagi:2014qza}.
Similarly, they may be used to infer the equation of state (EOS) during
the quark-hadron phase transition in quantum
chromodynamics~\cite{Schettler:2010dp,Hajkarim:2019csy} or
constrain properties of hidden sectors beyond the
Standard Model (BSM)~\cite{Jinno:2012xb,Caldwell:2018giq}.
The same may also be true for GWs emitted by cosmic
strings~\cite{Cui:2017ufi,Cui:2018rwi}, whose detailed
properties are currently under discussion
(see~\cite{Matsui:2019obe} and references therein).


The main goal of this paper is to demonstrate how future
observations of the primordial SGWB from inflation
across a large range of frequencies will be instrumental
in constraining BSM physics.
To this end, we shall consider a broad class of new physics
models in which the presence of a scalar field $\phi$ in
the early Universe leads to a modified expansion history.
Such a situation can occur in a variety of BSM scenarios.
Possible BSM scalar fields that $\phi$ may be identified
as include, but are not limited to:
the Polonyi field in models of gravity-mediated
supersymmetry breaking~\cite{Coughlan:1983ci,Ellis:1986zt},
a modulus field in four-dimensional compactifications of string
theory~\cite{deCarlos:1993wie,Moroi:1999zb,Durrer:2011bi,Kane:2015jia},
a flavon field in extensions of the Standard Model (SM) that aim
at explaining the flavor structure of quark and leptons
(such as, \textit{e.g.}, the Froggatt-Nielsen
flavor model~\cite{Froggatt:1978nt}),
and the saxion field in supersymmetric axion
models~\cite{Kugo:1983ma,Banks:2002sd,Kawasaki:2008jc,
Harigaya:2013vja,Harigaya:2015soa,Co:2016vsi,Co:2017orl}.


At some point prior to BBN, the energy density stored in the coherent
oscillations of the scalar field begins to dominate over radiation.
The ensuing era of scalar-field domination (SD), or \textit{scalar era} for
short, leads to a characteristic feature in the transfer function
for primordial GWs, which may be detected by future GW experiments.
In a detailed numerical analysis, we calculate the transfer function
for primordial GWs that captures the modified expansion during the
scalar era.
This allows us to determine the final GW spectrum across the entire
relevant parameter space of our scalar-field model.
For each parameter point, we compute the signal-to-noise 
ratios (SNRs) at which the GW spectrum as well as individual features
in this spectrum are expected to be observed by various future GW experiments.
The outcome of our analysis is a global picture of the GW signature
of the scalar era and the experimental prospects of detecting it.
Our study highlights the complementarity of future GW experiments
across the entire range of accessible frequencies and defines an important
benchmark scenario for the experimental GW community in the coming decades.


\noindent\textbf{Primordial gravitational waves.} GWs correspond to
spatial tensor perturbations $h_{ij}$ of the spacetime metric,
\begin{align}
ds^2  = - dt^2 + a^2\left(t\right)
\left[\delta_{ij} + h_{ij}\left(t,\mathbf{x}\right)\right] dx^i dx^j \,,
\end{align}
where $a$ denotes the cosmic scale factor and the symmetric tensor
$h_{ij}$ is evaluated in transverse-traceless gauge,
$h_{ii} = \partial_i\,h_{ij} = 0$.
The free propagation of GWs is governed by the linearized Einstein equation
with a vanishing source term.
In Fourier space, one thus obtains the following equations of motion
for the tensor modes $h_{\mathbf{k}}^p$,
\begin{align}
\label{eq:hijEOM}
\left[\frac{d^2}{du^2} + \frac{2}{a\left(u\right)}\frac{da\left(u\right)}{du}\frac{d}{du}
+ 1 \right] h_{\mathbf{k}}^p\left(u\right) = 0  \,.
\end{align}
Here, $\mathbf{k}$ denotes the comoving 3-momentum with absolute value
$k = \left\lVert\mathbf{k}\right\rVert$, $p = +,\times$ distinguishes
the two possible polarization states of a GW, and the variable $u = k\tau$ (with $\tau$
denoting conformal time, $dt = a \, d\tau$) is a measure for the spatial extent of a GW.
For $u \ll 1$ ($u \gg 1$), a mode is located far outside
(inside) the Hubble horizon.


An important property of inflationary GWs is that they are stretched
to super-horizon size during inflation. 
Outside the horizon, GWs remain frozen, until they re-enter the horizon
during the decelerating expansion after inflation.
This enables one to factorize the amplitude of a GW mode at the present time $\tau_0$, 
$h_{\mathbf{k}}^p\left(\tau_0\right) = \bar{h}_{\mathbf{k}}^p\,
\chi_k\left(\bar{\tau},\tau_0\right)$,
into an initial value at $\bar{\tau}$ determined by inflation,
$\bar{h}_{\mathbf{k}}^p = h_{\mathbf{k}}^p\left(\bar{\tau}\right)$,
and a transfer function  $\chi_k\left(\bar{\tau},\tau_0\right)$ 
that accounts for the redshift subsequent to horizon re-entry.
For any expansion history after inflation [specified by
the function $a\left(u\right)$],
the transfer function can be found by solving Eq.~\eqref{eq:hijEOM}
for $\chi_k$ as a function of $u$ with boundary conditions $\chi_k = 1$ and
$\partial_u \chi_k = 0$ at $u = 0$.
The same factorization also carries over to $\Omega_{\rm GW}\left(f\right)$,
\textit{i.e.}, the GW energy density per logarithmic frequency interval $d\left(\ln f\right)$
in units of the critical energy density of a flat universe.
For modes deep inside the horizon, one obtains the following GW spectrum at the present time,
\begin{align}
\label{eq:OGW}
\Omega_{\rm GW}^0\left(f\right) \simeq
\frac{1}{12}\frac{k^2}{a_0^2H_0^2}\,\mathcal{P}_t\left(k\right)
\left|\chi_k\right|^2 \,, \quad f = \frac{k}{2\pi\,a_0} \,,
\end{align}
where $a_0$ and $H_0 = 100\,h\,\textrm{km}/\textrm{s}/\textrm{Mpc}$ represent
the present values of the scale factor and the Hubble rate, respectively.
The primordial tensor spectrum $\mathcal{P}_t$ depends
on the initial conditions set by the inflationary period and is conventionally
parametrized by a simple power law around the CMB pivot scale,
$k_{\rm CMB} = 0.05\,\textrm{Mpc}^{-1}$,
\begin{align}
\label{eq:Pt}
\mathcal{P}_t = \frac{k^3}{\pi^2}\left(\left|\bar{h}_{\mathbf{k}}^+\right|^2
+ \left|\bar{h}_{\mathbf{k}}^\times\right|^2\right) =
A_t \left(\frac{k}{k_{\rm CMB}}\right)^{n_t} \,.
\end{align}


In this paper, we are interested in assessing the maximal reach
of future GW experiments.
For this reason, we decide to fix the primordial tensor amplitude
at the maximally allowed value, $A_t \simeq 1.5 \times 10^{-10}$,
that is consistent with the most recent upper bound on
the tensor-to-scalar ratio, $r = A_t / A_s \lesssim 0.07$~\cite{Akrami:2018odb},
where $A_s \simeq 2.1 \times 10^{-9}$ is the measured amplitude
of the primordial scalar spectrum.
Similarly, we shall assume a blue-tilted tensor spectrum, fixing its
index at an optimistic value of $n_t = 0.4$.
On the one hand, this value clearly violates the consistency relation
of standard single-field slow-roll inflation, $n_t = -r/8 < 0$.
This means that we implicitly assume a nonminimal inflationary sector that is
capable of generating a strong SGWB at small scales.
Such scenarios exist in the literature, with one
prominent example being natural inflation coupled to gauge
fields~\cite{Cook:2011hg,Barnaby:2011qe,Anber:2012du,Domcke:2016bkh,
Jimenez:2017cdr,Papageorgiou:2019ecb}.
On the other hand, our choice for the spectral index is
phenomenologically still completely viable.
$n_t = 0.4$ complies with (1) the CMB constraints derived by the PLANCK
Collaboration~\cite{Akrami:2018odb}, (2) the upper bound
on the amplitude of an isotropic SGWB at $\mathcal{O}\left(10\right)\,\textrm{Hz}$
frequencies by LIGO and Virgo~\cite{TheLIGOScientific:2016dpb,LIGOScientific:2019vic},
and (3) global bounds on primordial GWs (based, \textit{e.g.}, on BBN)
across the full frequency spectrum~\cite{Kuroyanagi:2014nba,Lasky:2015lej}.


\vspace{0.1cm}\noindent\textbf{The scalar era.} We consider the dynamics
of a real scalar field $\phi$, employing an effective
description in terms of three parameters: mass $m_\phi$,
decay rate $\Gamma_\phi$, and initial field value $\phi_{\rm ini}$
at the end of inflation.
We assume a harmonic potential around the origin in field
space and some (direct or indirect) coupling to the Standard Model
that allows $\phi$ to decay to radiation at the end of its lifetime.
The dynamical evolution of the field $\phi$ is
governed by the Klein-Gordon equation in an expanding background,
\begin{align}
\label{eq:phiEOM}
\left[\frac{d^2}{dt^2} + \left(3\,H\left(t\right) + \Gamma_\phi\right)
\frac{d}{dt} + m_\phi^2\right]\phi\left(t\right) = 0 \,,
\end{align}
where we included a friction term proportional to $\Gamma_\phi$.
The Hubble rate $H$ follows from the Friedmann equation,
\begin{align}
\label{eq:aEOM}
H^2\left(t\right) = \left[\frac{\dot{a}\left(t\right)}{a\left(t\right)}\right]^2
= \frac{\rho_\phi\left(t\right) + \rho_R\left(t\right)}{3\,M_{\rm Pl}^2} \,,
\end{align}
where $M_{\rm Pl} \simeq 2.44 \times 10^{18}\,\textrm{GeV}$ is
the reduced Planck mass and
$\rho_\phi = 1/2\,\dot{\phi}^2 + 1/2\,m_\phi^2\,\phi^2$ denotes the energy 
density stored in the field $\phi$.
The energy density of radiation, $\rho_R$, is described by an evolution equation that
follows from covariant energy conservation in an expanding universe,
\begin{align}
\label{eq:REOM}
\left[\frac{d}{dt} + 4\,\frac{g_{*,s}\left(\rho_R\left(t\right)\right)}
{g_{*,\rho}\left(\rho_R\left(t\right)\right)}\,
H\left(t\right)\right]\rho_R\left(t\right) = \Gamma_\phi\,\dot{\phi}^2\left(t\right)\,,
\end{align}
Here, $g_{*,s}$ and $g_{*,\rho}$ are the effective numbers of relativistic
degrees of freedom that contribute to the entropy and energy densities
of radiation, respectively.
We will evaluate  $g_{*,s}$ and $g_{*,\rho}$ using
the numerical data tabulated in~\cite{Saikawa:2018rcs}.


\begin{figure}
\begin{center}
\includegraphics[width=0.48\textwidth]{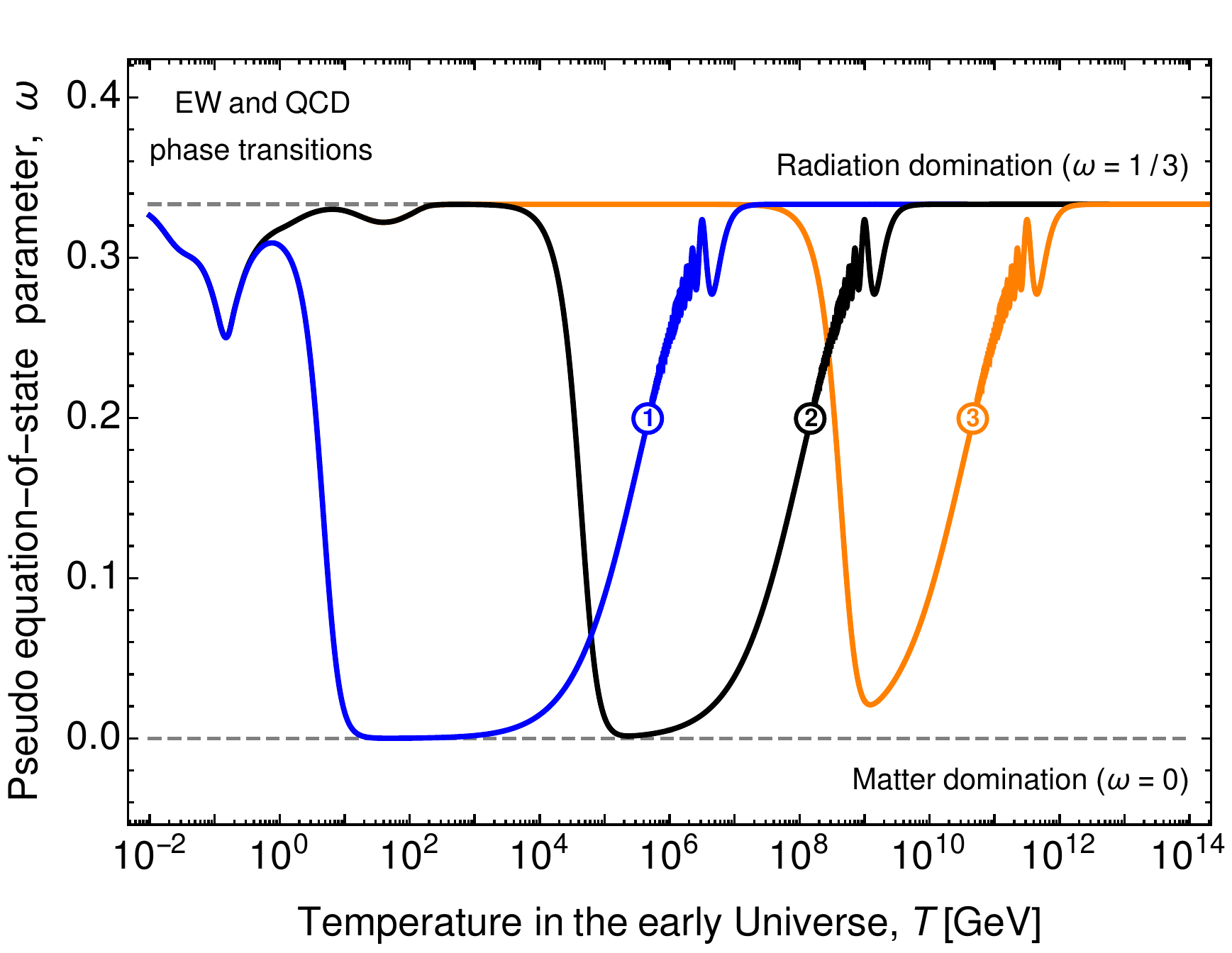} 
\caption{Pseudo EOS parameter $\omega = 2/\left(3\,tH\right)-1$ as a function
of temperature $T$ for three parameter examples (see Fig.~\ref{fig:scan}).
$\omega$ is defined such that $a \propto t^\alpha$,
where $\alpha = 2/3/\left(1+\omega\right)$.
It equals the EOS parameter $w = p/\rho$ as long as $w = \textrm{const}$.}
\label{fig:omega}
\end{center}
\end{figure}


Let us now discuss the system of equations~\eqref{eq:phiEOM} to \eqref{eq:REOM}.
First, we observe that $\phi$ only begins to roll\,---\,and eventually
oscillate around the origin in field space\,---\,once
the Hubble rate has dropped below the scalar mass, $H \lesssim m_\phi$.
At earlier times, the Hubble friction term in~\eqref{eq:phiEOM}
keeps $\phi$ more or less fixed at $\phi_{\rm ini}$.
For a sufficiently high reheating temperature after inflation,
the expansion is initially driven by radiation.
$\rho_R$ is, however, rapidly redshifted,
$\rho_R \propto 1/a^4$, such that, at some temperature $T_{\rm SD}$,
the energy density $\rho_\phi$ begins to dominate the expansion,
\begin{align}
\label{eq:TSD}
T_{\rm SD} \simeq
\frac{b^2}{\alpha_{\rm SD}^{3/2}} \left(\frac{5}
{36\pi^2\,g_{*,\rho}^{\rm SD}}\right)^{1/4}
\left(\frac{\phi_{\rm ini}}{M_{\rm Pl}}\right)^2\left(m_\phi\,M_{\rm Pl}\right)^{1/2} .
\end{align}
Here, $b = \Gamma\left(5/4\right)/\Gamma\left(3/2\right) \simeq 1.02$
stems from the analytic solution of~\eqref{eq:phiEOM} in a radiation-dominated (RD)
background, while $\alpha_{\rm SD} \approx 2/3$ characterizes the behavior
of the scale factor at the onset of scalar-field domination,
$a\left(t\right) \propto t^{\alpha_{\rm SD}}$. 
The result in~\eqref{eq:TSD} is only valid for
$\phi_{\rm ini} \lesssim M_{\rm Pl}$.
Otherwise, the scalar field has not yet begun to
oscillate at the onset of the scalar era, which leads to a short period
of vacuum domination.
Such a second phase of inflation results in a nontrivial
oscillatory feature in the transfer function~\cite{Jinno:2013xqa}.
On the other hand, it also causes a stronger dilution of the primordial
SGWB, which weakens the expected GW signal.
For this reason, we will not consider the case
$\phi_{\rm ini} \gtrsim M_{\rm Pl}$ in this work. 


After many oscillations around the potential minimum, $t \gg 1/m_\phi$,
the energy density of the scalar field behaves like the energy
density of pressureless dust, $\rho_\phi \propto 1/a^3$.
The scalar era therefore effectively mimics, at least
for some intermediate period, a phase of
matter-dominated expansion (see Fig.~\ref{fig:omega}).
It lasts until $\phi$ decays into SM radiation around
$t_{\rm RD} \simeq c/\Gamma_\phi$, which marks the onset 
of the final radiation-dominated era at temperature $T_{\rm RD}$,
\begin{align}
T_{\rm RD} = \bigg(\frac{\alpha_{\rm RD}}{c}\bigg)^{1/2}
\left(\frac{90}{2\pi^2\,g_{*,\rho}^{\rm RD}}\right)^{1/4}
\left(\Gamma_\phi\,M_{\rm Pl}\right)^{1/2} \,,
\end{align}
where $\alpha_{\rm RD} \approx 2/3$.
Assuming a constant EOS throughout the scalar era,
one is able to analytically derive $c \simeq 1.07$.


The nonstandard EOS and the late-time entropy injection during the scalar era
result in the dilution of sub-horizon tensor modes. This effect is captured
by the dilution factor $D = \left[S\left(T \gg T_{\rm SD}\right)/
S\left(T \ll T_{\rm RD}\right)\right]^{1/3}$, where $S$ denotes
the comoving entropy density, $S = a^3\,s$,
\begin{align}
D = e^{c/3}\,d\,
\bigg(\frac{g_{*,s}^{\rm SD}}{g_{*,s}^{\rm RD}}\bigg)^{1/3}
\bigg(\frac{g_{*,\rho}^{\rm RD}}{g_{*,\rho}^{\rm SD}}\bigg)^{1/4}
\eta^{1/6} \,.
\end{align}
The factor $d \simeq 0.80$, which needs to be determined numerically,
accounts for entropy production at $T < T_{\rm RD}$.
The factor $\eta$ is defined as the ratio of the two values
of the Hubble rate at $T = T_{\rm RD}$ and $T = T_{\rm SD}$,
respectively,
\begin{align}
\label{eq:eta}
\eta = \frac{H_{\rm RD}}{H_{\rm SD}} = 
\frac{18\,\alpha_{\rm SD}^3\,\alpha_{\rm RD}}{b^4\,c}
\frac{\Gamma_\phi}{m_\phi}
\left(\frac{M_{\rm Pl}}{\phi_{\rm ini}}\right)^4 \,.
\end{align}


The two temperatures $T_{\rm SD}$ and $T_{\rm RD}$ also single out two frequencies
in the GW spectrum, $f_{\rm SD}$ and $f_{\rm RD}$, which correspond to the
tensor modes that re-enter the horizon at these temperatures,
respectively.
For $f_{\rm RD}$, we find
\begin{align}
\label{eq:fRD}
\frac{f_{\rm RD}}{f_0} = d\,
\bigg(\frac{g_{*,s}^0}{g_{*,s}^{\rm RD}}\bigg)^{1/3}
\bigg(\frac{g_{*,\rho}^{\rm RD}}{g_{*,\rho}^0}\bigg)^{1/2}
\left(\frac{\Omega_R^0}{1/2}\right)^{1/2} \frac{T_{\rm RD}}{T_0} \,,
\end{align}
where $T_0 \simeq 2.72548\,\textrm{K} \simeq 2.3 \times 10^{-13}\,\textrm{GeV}$
is the current temperature of the CMB~\cite{Fixsen:2009ug},
$h^2\Omega_R^0 \simeq 4.2 \times 10^{-5}$ denotes
the energy density parameter of radiation at the present time,
and $f_0 = H_0/\left(2\pi\right)$ corresponds to the frequency of 
the GW mode that currently stretches across one Hubble radius,
$f_0/h \simeq 5.2 \times 10^{-19}\,\textrm{Hz}$.
The frequency $f_{\rm SD}$ can be computed in terms
of the quantities in \eqref{eq:eta} and \eqref{eq:fRD},
\begin{align}
f_{\rm SD} = \left(\frac{e^c}{\eta}\right)^{1/3} f_{\rm RD} \,.
\end{align}
This shows that the parameter $\eta$ determines
the duration of the scalar era in frequency space.
In our numerical analysis, we will scan over different values
of $\eta$ by varying the ratio $m_\phi/\Gamma_\phi$, while keeping
the initial field value fixed at $\phi_{\rm ini} = 10^{18}\,\textrm{GeV}$.
This benchmark represents a typical value that can be realized in many models as an initial condition after inflation~\cite{Bunch:1978yq,Linde:1982uu,Affleck:1984fy,Starobinsky:1994bd}.
Aside from that, it is important to note that, to first approximation, it is straightforward to generalize our analysis to different values of $\phi_{\rm ini}$ by simply rescaling all values of $m_\phi/\Gamma_\phi$. 

\begin{figure}
\begin{center}
\includegraphics[width=0.48\textwidth]{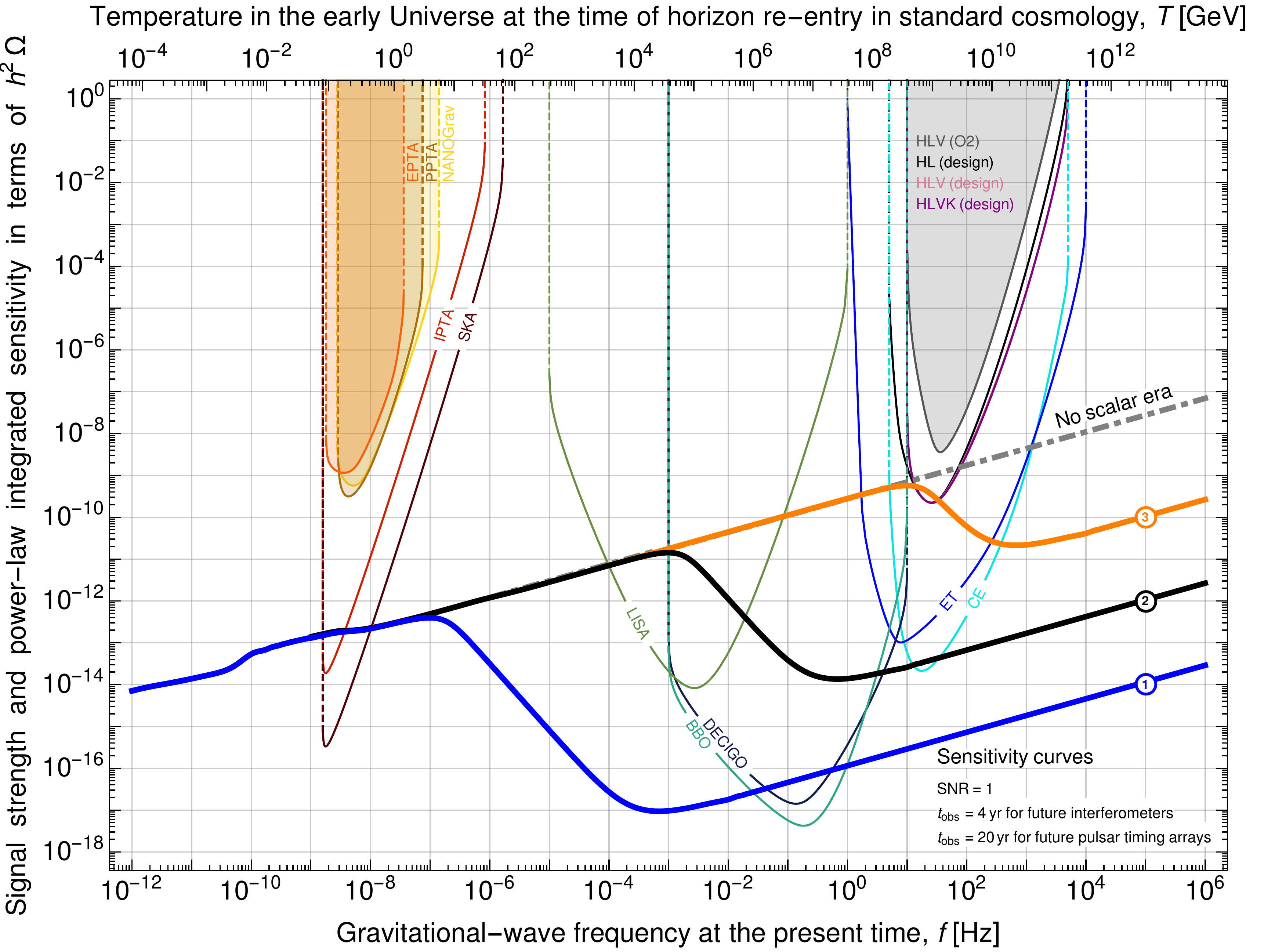} 
\caption{Final GW spectrum for three parameter examples (see Fig.~\ref{fig:scan})
alongside the power-law integrated sensitivity curves~\cite{Thrane:2013oya}
of existing and upcoming GW experiments.}
\label{fig:spectrum}
\end{center}
\end{figure}

\vspace{0.1cm}\noindent\textbf{Transfer function.} For a mode with wave number $k$
deep inside the horizon, the transfer function roughly scales
like $\chi_k \sim a_k/a_0$, where $a_k$ is the scale factor at the
time of horizon re-entry, implicitly defined
by $a_k = k / H\left(a_k\right)$~\cite{Turner:1993vb,Boyle:2005se,Watanabe:2006qe}.
For an EOS $w = p/\rho = \textrm{const}$ (with pressure $p$
and energy density $\rho$), this results
in a simple power-law scaling of the final GW spectrum,
\begin{align}
\Omega_{\rm GW}^0\left(f\right) \propto f^n \,, \quad
n = n_t + \frac{2\left(3\,w-1\right)}{3\,w+1} \,,
\end{align}
which illustrates the main impact of the scalar era on the GW spectrum:
For frequencies $f_{\rm RD} \lesssim f \lesssim f_{\rm SD}$, the
spectral index $n$ changes by $\Delta n \simeq -2$ compared to the spectrum
at smaller and larger frequencies.
This results in a step-like feature in the GW spectrum with two
kinks located at $f \simeq f_{\rm RD}$ and $f \simeq f_{\rm SD}$,
respectively (see Fig.~\ref{fig:spectrum}).
The ratio of the spectral amplitudes at these two frequencies is
controlled by the dilution factor $D$. 
Neglecting any changes in $g_{*,\rho}$ and $g_{*,s}$, we approximately
find
\begin{align}
\frac{\Omega_{\rm GW}^0\left(f_{\rm SD}\right)}{\Omega_{\rm GW}^0\left(f_{\rm RD}\right)}
\simeq \left(\frac{f_{\rm SD}}{f_{\rm RD}}\right)^{n_t} D^4 \,.
\end{align}


To obtain a precise and accurate result for the transfer function, we numerically solve
the coupled set of equations~\eqref{eq:hijEOM}, \eqref{eq:phiEOM},
\eqref{eq:aEOM}, and \eqref{eq:REOM} for any choice of
$m_\phi$ and $\Gamma_\phi$.
Together with \eqref{eq:OGW} and \eqref{eq:Pt}, this allows us to compute the final
GW spectrum across our entire parameter space.


\vspace{0.1cm}\noindent \textbf{Signal-to-noise ratios.}  A GW experiment
with strain noise power spectrum
$S_{\rm noise} = 6 f_0^2/f^3 \Omega_{\rm noise}$ is able
to detect a stochastic signal $\Omega_{\rm signal}$ at the following
SNR~\cite{Allen:1997ad},
\begin{align}
\label{eq:snr}
\textrm{SNR}^2 = N\,t_{\rm obs}
\int_{f_{\rm min}}^{f_{\rm max}}df
\left[\frac{\Omega_{\rm signal}\left(f\right)}{\Omega_{\rm noise}\left(f\right)}
\right]^2 \,,
\end{align}
where $t_{\rm obs}$ denotes the experiment's observing time and $N=1$ ($N=2$)
for experiments that perform an auto (cross) correlation measurement of the SGWB.
With the full GW spectrum at hand, we are now able to forecast
the SNRs for future ground-based interferometers
[(1) the four-detector network (\textbf{HLVK}) consisting of advanced LIGO in
Hanford and Livingston~\cite{Harry:2010zz,TheLIGOScientific:2014jea}, advanced
Virgo~\cite{TheVirgo:2014hva}, and KAGRA~\cite{Somiya:2011np,Aso:2013eba},
(2) Cosmic Explorer (\textbf{CE})~\cite{Evans:2016mbw}, and
(3) Einstein Telescope (\textbf{ET})~\cite{Punturo:2010zz,Hild:2010id,Sathyaprakash:2012jk}],
future spaced-based interferometers
[(1) \textbf{LISA}~\cite{Audley:2017drz}, (2) \textbf{DECIGO}~\cite{Seto:2001qf,Kawamura:2006up},
and (3) \textbf{BBO}~\cite{Crowder:2005nr,Corbin:2005ny,Harry:2006fi}], and future
pulsar timing arrays [(1) \textbf{IPTA}~\cite{Hobbs:2009yy,IPTA):2013lea,Verbiest:2016vem} and
(2) \textbf{SKA}~\cite{Carilli:2004nx,Janssen:2014dka,Bull:2018lat}].
We assume an observing time $t_{\rm obs} = 4\,\textrm{yr}$ for
all interferometer experiments and the equivalent~\cite{Siemens:2013zla} of
$t_{\rm obs} = 20\,\textrm{yr}$ for IPTA and SKA.
In future work, one might consider refining our analysis by means of a Fisher information analysis.
We, however, expect that such an analysis would only marginally improve over our SNR-based approach (see~\cite{Caldwell:2018giq}, which finds no difference between both approaches in the specific case of LISA).


In principle, it is also necessary to account for confusion noise from
astrophysical sources (see, \textit{e.g.},~\cite{Cornish:2018dyw}).
The significance of confusion noise, however, decreases over time,
as larger observing times allow one to resolve and subtract a larger
number of individual foreground sources.
In what follows, we will therefore consider an idealized scenario
without any foreground contamination.
In this sense, our SNR forecasts may also be regarded
as upper limits on what will be realistically achievable.

\begin{figure}
\begin{center}
\includegraphics[width=0.48\textwidth]{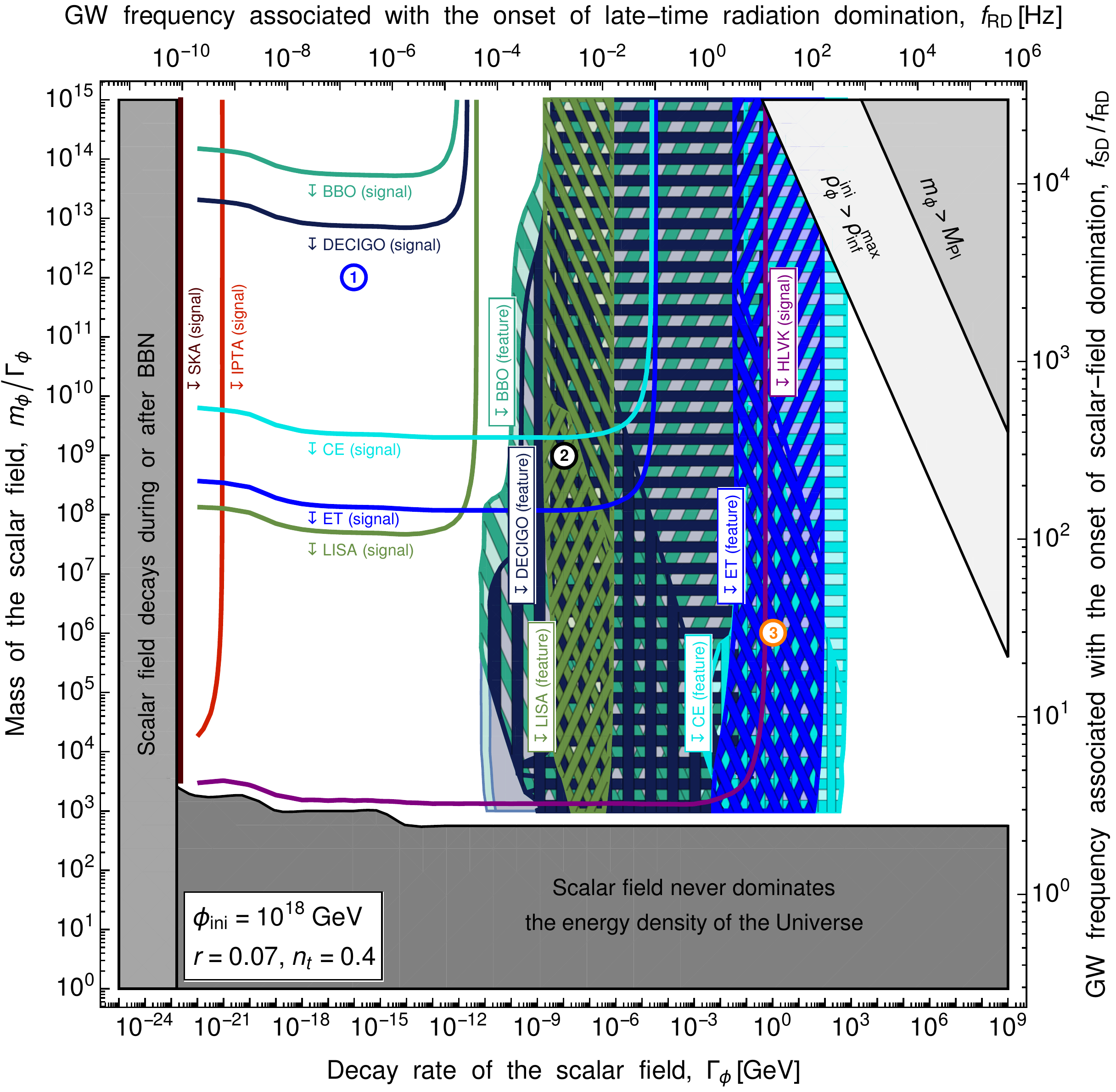} 
\caption{Experimental prospects to observe a \textit{signal} or a even
a \textit{feature} in the GW spectrum (see text for details)
in dependence of the model parameters $m_\phi$ and $\Gamma_\phi$.
The three points \ding{192}, \ding{193}, \ding{194} mark the parameter examples
used in Figs.~\ref{fig:omega} and \ref{fig:spectrum}.
The meshes drawn in the \textit{feature} regions indicate
where the frequencies $f_{\rm RD}$ (meshes extending up to arbitrarily large $m_\phi$)
and $f_{\rm SD}$ (meshes extending up to a finite $m_\phi$) fall into
the frequency ranges of the individual experiments (see Fig.~\ref{fig:spectrum}).}
\label{fig:scan}
\end{center}
\end{figure}

For each experiment, we compute two kinds of SNRs (see Fig.~\ref{fig:scan}):
First, we compute the total SNR based on the full GW signal,
$\Omega_{\rm signal} \rightarrow \Omega_{\rm GW}^0$.
If this SNR is larger than 1, we conclude that the respective experiment 
will have ``at least a slight chance'' of detecting a \textit{signal}.
Then, we compute a second, reduced SNR based on a subtracted GW signal,
$\Omega_{\rm signal} \rightarrow \Omega_{\rm GW}^0 - \Omega_{\rm GW}^{\rm fit}$,
where $\Omega_{\rm GW}^{\rm fit}$ represents a power-law fit of the GW spectrum
around the frequency at which a given experiment is most sensitive.
This construction is in accord with the matched-filter approaches in~\cite{Caldwell:2018giq,Kuroyanagi:2018csn}.
If the reduced SNR is larger than 100, we conclude that the respective experiment 
is ``very likely'' to detect a \textit{feature} in the SGWB,
\textit{i.e.}, a deviation from a pure power law, related to the scalar era.
In practice, this feature will correspond to one or both of the two
kinks at $f \simeq f_{\rm RD}$ and $f \simeq f_{\rm SD}$.
Our choice of SNR threshold values allows us to assign the labels ``at least a slight chance'' and ``very likely'' to the different regions in Fig.~\ref{fig:scan} in a conservative manner.
Meanwhile, we stress that our quantitative conclusions are nearly insensitive to $\mathcal{O}\left(1\right)$ changes of these threshold values.
Our qualitative conclusions are not affected at all by the precise values of the SNR thresholds.


\vspace{0.1cm}\noindent\textbf{Discussion.}  Our results
in Fig.~\ref{fig:scan} are a powerful testament to the
complementarity of future GW experiments.
Consider, \textit{e.g.}, a scalar field with $m_\phi = 10\,\textrm{GeV}$,
$\Gamma_\phi = 10\,\textrm{eV}$, and $\phi_{\rm ini} = 10^{18}\,\textrm{GeV}$
(point \ding{193} in Fig.~\ref{fig:scan}).
In this case, the scalar era occurs at temperatures
$5.4\times10^4 \lesssim T/\textrm{GeV}\lesssim 1.7\times10^8$, which
leads to a dilution factor $D \simeq 7.9 \times 10^{-2}$ and
a modification of the GW spectrum at frequencies
$1.6\times10^{-3}\lesssim f/\textrm{Hz}\lesssim 4.9\times10^{-1}$.
This scenario can be directly probed by LISA, DECIGO, and BBO, all of
which will observe a departure from a pure power law.
At the same time, CE, IPTA, and SKA will detect the SGWB from inflation,
while HLVK and ET will fail to observe a primordial signal.
On top of that, future CMB observations may confirm
the blue tilt of the tensor spectrum.
Along these lines, it is possible to translate every
point in Fig.~\ref{fig:scan} into a list of
predictions, \textit{i.e.}, a characteristic experimental fingerprint,
for the GW experiments that we have considered in this paper.


The results of our analysis are applicable to a wide range
of particle physics models that lead to a scalar era in the early Universe.
In addition, it is important to note that the modified
expansion history and late-time entropy production during the scalar
era not only affect the spectrum of primordial GWs,
but also the evolution of other cosmological relics,
such as dark matter~\cite{Hashimoto:1998ua,Giudice:2000ex,Acharya:2009zt,
Monteux:2015qqa,Co:2015pka,Roszkowski:2015psa,Hamdan:2017psw,Nelson:2018via}
and the baryon asymmetry of the Universe~\cite{Chen:2019wnk}.
Finding evidence for a scalar era in future GW experiments would
therefore have a profound impact on our understanding of
BSM physics and early-Universe cosmology alike.


The analysis presented in this paper should only be regarded as
a first step.
Next, it will be necessary to relax our assumptions
regarding the primordial tensor spectrum, the initial field
value, and the shape of the scalar potential.
Aside from that, it would be interesting to identify a self-consistent
embedding of our scenario in a concrete inflationary model
that is capable of generating
a strongly blue-tilted tensor spectrum.
We leave these steps for future work,
concluding with the observation
that a scalar era in the early Universe represents a
fascinating possibility that awaits further exploration.


\vspace{0.1cm}\textit{Acknowledgements.} We thank Sachiko Kuroyanagi for
sharing with us her numerical results on the DECIGO overlap reduction function;
Ken'ichi Saikawa and Satoshi Shirai for
sharing with us their numerical results on the GW
transfer function at frequencies below $1\,\textrm{nHz}$, which takes
into account damping effects caused by free-streaming photons
and neutrinos at low temperatures;
and Hirotaka Yuzurihara for useful comments on the
KAGRA sensitivity curve.
We are grateful to all of these authors as well as to Kohei Kamada,
Marco Peloso, and Angelo Ricciardone for helpful discussions.
This project has received funding/support from the European Union's
Horizon 2020 research and innovation programme under the
Marie Sk\l odowska-Curie grant agreement No.\ 690575.
This work was supported by INFN
through the ``Theoretical Astroparticle Physics'' (TAsP) project.


\bibliographystyle{JHEP}
\bibliography{arxiv_3}


\end{document}